\documentstyle[12pt]{article}
\advance \topmargin by -\headheight
\advance \topmargin by -\headsep

\evensidemargin \oddsidemargin
\begin{document}

\topmargin -.6in
\def\rf#1{(\ref{eq:#1})}
\def\lab#1{\label{eq:#1}}
\def\nonu{\nonumber}
\def\br{\begin{eqnarray}}
\def\er{\end{eqnarray}}
\def\be{\begin{equation}}
\def\ee{\end{equation}}
\def\eq{\!\!\!\! &=& \!\!\!\! }
\def\foot#1{\footnotemark\footnotetext{#1}}
\def\lb{\lbrack}
\def\rb{\rbrack}
\def\llangle{\left\langle}
\def\rrangle{\right\rangle}
\def\blangle{\Bigl\langle}
\def\brangle{\Bigr\rangle}
\def\llbrack{\left\lbrack}
\def\rrbrack{\right\rbrack}
\def\lcurl{\left\{}
\def\rcurl{\right\}}
\def\({\left(}
\def\){\right)}
\newcommand{\nit}{\noindent}
\newcommand{\ct}[1]{\cite{#1}}
\newcommand{\bi}[1]{\bibitem{#1}}
\def\lskip{\vskip\baselineskip\vskip-\parskip\noindent}
\relax

\def\tr{\mathop{\rm tr}}
\def\Tr{\mathop{\rm Tr}}
\def\v{\vert}
\def\bv{\bigm\vert}
\def\Bgv{\;\Bigg\vert}
\def\bgv{\bigg\vert}
\newcommand\partder[2]{{{\partial {#1}}\over{\partial {#2}}}}
\newcommand\funcder[2]{{{\delta {#1}}\over{\delta {#2}}}}
\newcommand\Bil[2]{\Bigl\langle {#1} \Bigg\vert {#2} \Bigr\rangle}  
\newcommand\bil[2]{\left\langle {#1} \bigg\vert {#2} \right\rangle} 
\newcommand\me[2]{\left\langle {#1}\right|\left. {#2} \right\rangle} 
\newcommand\sbr[2]{\left\lbrack\,{#1}\, ,\,{#2}\,\right\rbrack}
\newcommand\pbr[2]{\{\,{#1}\, ,\,{#2}\,\}}
\newcommand\pbbr[2]{\lcurl\,{#1}\, ,\,{#2}\,\rcurl}
%
\def\a{\alpha}
\def\b{\beta}
\def\dc{{\cal D}}
\def\d{\delta}
\def\D{\Delta}
\def\eps{\epsilon}
\def\vareps{\varepsilon}
\def\g{\gamma}
\def\G{\Gamma}
\def\grad{\nabla}
\def\h{{1\over 2}}
\def\l{\lambda}
\def\L{\Lambda}
\def\m{\mu}
\def\n{\nu}
\def\o{\over}
\def\om{\omega}
\def\O{\Omega}
\def\p{\phi}
\def\P{\Phi}
\def\pa{\partial}
\def\pr{\prime}
\def\ra{\rightarrow}
\def\s{\sigma}
\def\S{\Sigma}
\def\t{\tau}
\def\th{\theta}
\def\Th{\Theta}
\def\ti{\tilde}
\def\wti{\widetilde}
\def\jc{J^C}
\def\bj{{\bar J}}
\def\sj{{\jmath}}
\def\bsj{{\bar \jmath}}
\def\bp{{\bar \p}}
\def\faa{Fa\'a di Bruno~}
\def\ca{{\cal A}}
\def\cb{{\cal B}}
\def\ce{{\cal E}}
\newcommand\sumi[1]{\sum_{#1}^{\infty}}   
\newcommand\fourmat[4]{\left(\begin{array}{cc}  
{#1} & {#2} \\ {#3} & {#4} \end{array} \right)}

%
\def\lie{{\cal G}}
\def\dlie{{\cal G}^{\ast}}
\def\elie{{\widetilde \lie}}
\def\edlie{{\elie}^{\ast}}
\def\hlie{{\cal H}}
\def\wlie{{\widetilde \lie}}
\def\f#1#2#3 {f^{#1#2}_{#3}}
\def\winf{{\sf w_\infty}}
\def\win1{{\sf w_{1+\infty}}}
\def\hwinf{{\sf {\hat w}_{\infty}}}
\def\Winf{{\sf W_\infty}}
\def\Win1{{\sf W_{1+\infty}}}
\def\hWinf{{\sf {\hat W}_{\infty}}}
\def\Rm#1#2{r(\vec{#1},\vec{#2})}          
\def\OR#1{{\cal O}(R_{#1})}           
\def\ORti{{\cal O}({\widetilde R})}           
\def\AdR#1{Ad_{R_{#1}}}              
\def\dAdR#1{Ad_{R_{#1}^{\ast}}}      
\def\adR#1{ad_{R_{#1}^{\ast}}}       
\def\KP{${\rm \, KP\,}$}                 
\def\KPl{${\rm \,KP}_{\ell}\,$}         
\def\KPo{${\rm \,KP}_{\ell = 0}\,$}         
\def\mKPa{${\rm \,KP}_{\ell = 1}\,$}    
\def\mKPb{${\rm \,KP}_{\ell = 2}\,$}    
%
\def\rlx{\relax\leavevmode}
\def\inbar{\vrule height1.5ex width.4pt depth0pt}
\def\IZ{\rlx\hbox{\sf Z\kern-.4em Z}}
\def\IR{\rlx\hbox{\rm I\kern-.18em R}}
\def\IC{\rlx\hbox{\,$\inbar\kern-.3em{\rm C}$}}
\def\one{\hbox{{1}\kern-.25em\hbox{l}}}
\def\0#1{\relax\ifmmode\mathaccent"7017{#1}%
        \else\accent23#1\relax\fi}
\def\omz{\0 \omega}
%
\def\ltimes{\mathrel{\vrule height1ex}\joinrel\mathrel\times}
\def\rtimes{\mathrel\times\joinrel\mathrel{\vrule height1ex}}
%
\def\mark{\noindent{\bf Remark.}\quad}
\def\prop{\noindent{\bf Proposition.}\quad}
\def\theor{\noindent{\bf Theorem.}\quad}
\def\name{\noindent{\bf Definition.}\quad}
\def\exam{\noindent{\bf Example.}\quad}
\def\proof{\noindent{\bf Proof.}\quad}
%
%
\def\PRL#1#2#3{{\sl Phys. Rev. Lett.} {\bf#1} (#2) #3}
\def\NPB#1#2#3{{\sl Nucl. Phys.} {\bf B#1} (#2) #3}
\def\NPBFS#1#2#3#4{{\sl Nucl. Phys.} {\bf B#2} [FS#1] (#3) #4}
\def\CMP#1#2#3{{\sl Commun. Math. Phys.} {\bf #1} (#2) #3}
\def\PRD#1#2#3{{\sl Phys. Rev.} {\bf D#1} (#2) #3}
\def\PLA#1#2#3{{\sl Phys. Lett.} {\bf #1A} (#2) #3}
\def\PLB#1#2#3{{\sl Phys. Lett.} {\bf #1B} (#2) #3}
\def\JMP#1#2#3{{\sl J. Math. Phys.} {\bf #1} (#2) #3}
\def\PTP#1#2#3{{\sl Prog. Theor. Phys.} {\bf #1} (#2) #3}
\def\SPTP#1#2#3{{\sl Suppl. Prog. Theor. Phys.} {\bf #1} (#2) #3}
\def\AoP#1#2#3{{\sl Ann. of Phys.} {\bf #1} (#2) #3}
\def\PNAS#1#2#3{{\sl Proc. Natl. Acad. Sci. USA} {\bf #1} (#2) #3}
\def\RMP#1#2#3{{\sl Rev. Mod. Phys.} {\bf #1} (#2) #3}
\def\PR#1#2#3{{\sl Phys. Reports} {\bf #1} (#2) #3}
\def\AoM#1#2#3{{\sl Ann. of Math.} {\bf #1} (#2) #3}
\def\UMN#1#2#3{{\sl Usp. Mat. Nauk} {\bf #1} (#2) #3}
\def\FAP#1#2#3{{\sl Funkt. Anal. Prilozheniya} {\bf #1} (#2) #3}
\def\FAaIA#1#2#3{{\sl Functional Analysis and Its Application} {\bf #1} (#2)
#3}
\def\BAMS#1#2#3{{\sl Bull. Am. Math. Soc.} {\bf #1} (#2) #3}
\def\TAMS#1#2#3{{\sl Trans. Am. Math. Soc.} {\bf #1} (#2) #3}
\def\InvM#1#2#3{{\sl Invent. Math.} {\bf #1} (#2) #3}
\def\LMP#1#2#3{{\sl Letters in Math. Phys.} {\bf #1} (#2) #3}
\def\IJMPA#1#2#3{{\sl Int. J. Mod. Phys.} {\bf A#1} (#2) #3}
\def\AdM#1#2#3{{\sl Advances in Math.} {\bf #1} (#2) #3}
\def\RMaP#1#2#3{{\sl Reports on Math. Phys.} {\bf #1} (#2) #3}
\def\IJM#1#2#3{{\sl Ill. J. Math.} {\bf #1} (#2) #3}
\def\APP#1#2#3{{\sl Acta Phys. Polon.} {\bf #1} (#2) #3}
\def\TMP#1#2#3{{\sl Theor. Mat. Phys.} {\bf #1} (#2) #3}
\def\JPA#1#2#3{{\sl J. Physics} {\bf A#1} (#2) #3}
\def\JSM#1#2#3{{\sl J. Soviet Math.} {\bf #1} (#2) #3}
\def\MPLA#1#2#3{{\sl Mod. Phys. Lett.} {\bf A#1} (#2) #3}
\def\JETP#1#2#3{{\sl Sov. Phys. JETP} {\bf #1} (#2) #3}
\def\JETPL#1#2#3{{\sl  Sov. Phys. JETP Lett.} {\bf #1} (#2) #3}
\def\PHSA#1#2#3{{\sl Physica} {\bf A#1} (#2) #3}
\def\PHSD#1#2#3{{\sl Physica} {\bf D#1} (#2) #3}
\begin{titlepage}
\vspace*{-1cm}
\noindent
July, 1993 \hfill{IFT-P.038/93}\\
\phantom{bla}
\hfill{hep-th/9307064}
\\
\vskip .3in

\begin{center}

{\large\bf On Non-linear W-Infinity Symmetry of }
\end{center}
\begin{center}
{\large\bf Generalized Liouville and Conformal Toda Models}
\end{center}
\normalsize
\vskip .4in

\begin{center}
{ H. Aratyn\footnotemark
\footnotetext{Work supported in part by U.S. Department of Energy,
contract DE-FG02-84ER40173 and by NSF, grant no. INT-9015799}}

\par \vskip .1in \noindent
Department of Physics \\
University of Illinois at Chicago\\
845 W. Taylor St.\\
Chicago, Illinois 60607-7059\\
\par \vskip .3in

\end{center}

\begin{center}
{L.A. Ferreira\footnotemark
\footnotetext{Work supported in part by CNPq}}, J.F. Gomes$^{\,2}$,
and A.H. Zimerman$^{\,2}$

\par \vskip .1in \noindent
Instituto de F\'{\i}sica Te\'{o}rica-UNESP\\
Rua Pamplona 145\\
01405-900 S\~{a}o Paulo, Brazil
\par \vskip .3in

\end{center}

\begin{center}
{\large {\bf ABSTRACT}}\\
\end{center}
\par \vskip .3in \noindent

Invariance under non-linear $\hWinf$ algebra is shown for the
two-boson Liouville type of model and its algebraic generalizations,
the extended conformal Toda models.
The realization of the corresponding generators in terms of two boson
currents within KP hierarchy is presented.

\end{titlepage}

\lskip
{\large {\bf 1. Introduction}}
\lskip
Recently there has been an increasing interest in applications of
${\sf W}$-infinity algebras in various areas of theoretical physics,
like strings, matrix models, Hall effect etc.
The ${\sf W}$-infinity symmetry appears in two related forms, the linear
$\Win1$ algebra, which arises as a deformation of the area preserving
symmetry and its non-linear deformation, so-called $\hWinf$
algebra.
These algebraic structures find their natural realization in the setting of
the first and second brackets of KP hierarchy \ct{yam91,DHP,YW92}.
A special realizations of KP hierarchy in terms of two bosons
have been proposed in literature \ct{YW9111,2boson}.
In this paper we work with realization of $\hWinf$ within this
context, using two Bose KP currents as building blocks of the
$\hWinf$ generators.

Despite of the increasing importance of ${\sf W}$-infinity algebras
the number of non-trivial (interacting) field theoretical models
displaying this type of symmetry remains very
limited.
The aim of this paper is to provide non-trivial invariant actions
under the non-linear $\hWinf$ symmetry.

One example of previously known model exhibiting ${\sf W}$-infinity
invariance is provided by the conformal affine Toda
(CAT) model \ct{2boson}.
The CAT model arises from the affine Toda model by introducing two extra
(ghost) fields, which makes it conformally invariant \ct{BB,AFGZ}.
These extra ghost fields bear responsibility for enlargement of symmetry.

In a similar manner one can extend the conformal symmetry of the Liouville
and Toda conformal actions by introducing ghost fields, which makes
spin-2 algebra anomaly free in such a way that a complete
action becomes invariant under non-linear $\hWinf$ algebra.
We call such models the generalized Liouville and Toda conformal
models. Extension of Liouville model of this type has been considered in
\ct{BI91} where ${\sf W}$-symmetry was discussed.

The main results of this paper are to be found in section 4, where we
provide Lagrangians of the generalized Liouville and Toda conformal
models. There we establish the precise form of $\hWinf$ field
transformations under which these models are invariant.

\lskip
{\large {\bf 2. Second Bracket Structure of KP Hierarchy}}
\lskip
We will study in this section the second-bracket structure within the
general KP setting with the Lax operator:
\be
L \equiv D +  \sum_{i=0}^{\infty} W_i (x, t) D^{-1-i}
\lab{laxop}
\ee
where $D = \pa / \pa x$ acts as an operator on quantities to the right of it.
Let us first introduce the second bracket structure as proposed by
Gelfand and Dickey (see e.g. \ct{dickey} for a review).
Define $X = \sum_{i=0}^{\infty} D^i x_i $ and the pairing
\be
\me {L} {X} \equiv \Tr \( L X\) = \int {\rm res} \( L X\)
= \int \sum_{k=0}^{\infty} W_k x_k
\lab{trace}
\ee
The Gelfand-Dickey second bracket structure is defined as
\be
\{ \, \me {L} {X} \, , \, \me {L} {Y} \, \}^{GD}_2 \equiv
\me {X} {L (YL)_{-} - (LY)_{-} L} \lab{gd2}
\ee

One obtains from \rf{gd2} by direct calculation
(see for instance \ct{DHP}):
\be
\{ W_n (x) \, , \, W_m (y) \}^{GD}_2 =  \O^{(1)}_{n,m}\(W (x)\)\,
\d (x-y) + \{ W_n (x) \, , \, W_m (y) \}^{GD}_2 \big\v_{\rm nonlinear}
\lab{gd2comp}
\ee
where the form $\O^{(r=1)}_{n,m}\(W(x)\)$ can be read from the general
expression:
\be
\O^{(r)}_{n,m}\(W(x)\) \equiv -\sum_{k=0}^{n+r} (-1)^k
{n+r \choose k} W_{n+m+r-k} (x) D^k_x +
\sum_{k=0}^{m+r} {m+r \choose k} D^k_x W_{n+m+r-k} (x)
\lab{wata}
\ee
This type of generalized forms \rf{wata}
reproduces the linear part of several higher brackets.
The nonlinear part in \rf{gd2comp} is given by
\br \lefteqn{
\{ W_n (x) \, , \, W_m (y) \}^{GD}_2 \big\v_{\rm nonlinear}
= \sum_{i=0}^{m-1} \Bigg\lb \sum_{k=1}^{m-i-1} {m-i-1 \choose k} W_i (x)
D^k_x W_{m+n-i-k-1} (x)        }\nonu \\
&-&\sum_{k=1}^{n} (-1)^k {n \choose k} W_{n+i-k} (x) D^k_x
W_{m-i-1} (x) \Bigg\rb \, \d (x-y)  \lab{nonligd2}\\
&-&\sum_{i=0}^{m-1} \sum_{k=0}^{n} \sum_{l =1}^{m-i-1} (-1)^k {n \choose k}
{m-i-1 \choose l} W_{n+i-k}(x) D^{k+l}_x W_{m-i-l-1} (x) \d (x-y) \nonu
\er
So far we had only defined Gelfand-Dickey bracket structures.
In order to
reproduce the Hamiltonian structure corresponding to KP hierarchy
flow equation one realizes that a further structure is required.
This structure can be obtained through the Dirac procedure
and is called a Drinfeld-Sokolov (DS)
bracket. It is given by
\be
\{ W_n (x) \, , \, W_m (y) \}^{DS} =
- \sum_{i=0}^{n-1} \sum_{j =0}^{m-1} (-1)^{n-i} {n \choose i}
{m \choose j} W_i (x) D^{n+m-i-j-1}_x W_j (x) \, \d (x-y) \lab{dscomp}
\ee
In conclusion the second Hamiltonian structure compatible with Lenard
relations is given by \ct{DHP}:
\be
\{ W_n (x) \, , \, W_m (y) \}_2 = \{ W_n (x) \, , \, W_m (y) \}_2^{GD}
+ \{ W_n (x) \, , \, W_m (y) \}^{DS}
\lab{2bracket}
\ee
We will now discuss realization of the second KP bracket structure in terms
of two related two-boson KP systems.
\lskip
{\large {\bf 3. Two-Boson KP Systems and their Miura Relations}}
\lskip
{\sl \faa Hierarchy.}
Consider truncated elements of KP hierarchy
of the type $L_{J} = D - J + \bj D^{-1} $, where we have introduced two
Bose currents $(\bj,J)$ \ct{2boson}.
It has been shown that such two-boson $(\bj,J)$ model is a consistent
restriction of the full KP hierarchy \ct{BAK85,ANPV}.
A calculation of the Poisson bracket
yields the first bracket structure  of two-boson $(\bj,J)$ system given
by $\pbr{J(x)}{\bj(y)}_1= - \d^{\pr} (x-y)$ and zero otherwise.

One easily associates the two-boson KP hierarchy with so called
\faa polynomials.
The trick is to introduce the gauge transformation
generated by $\P$ such that $\P^{\pr} = J$:
\be
L_{J}^{\pr} = e^{-\P} L_{J} e^{\P} = D + \bj \(D + J\)^{-1} =
D + \sumi{n=0} (-1)^n \bj P_n (J) D^{-1-n}
 \lab{faalax}
\ee
where $P_n (J)= (D + J)^n \cdot 1$ are the \faa polynomials.
As proven in \ct{ANPV} such gauge transformations as in \rf{faalax}
are symplectic and therefore we conclude that
\be
W_n = (-1)^n \bj P_n (J)
\lab{wnpj}
\ee
satisfy for the first bracket the Poisson-bracket structure of the linear
$\Win1\,$ algebra type, namely
${\pbr{W_n(x)}{W_m(x)}}_1 = \O^{(0)}_{nm} (W(x)) \d (x-y)$
described by the form $\O^{(0)}$ given in \rf{wata} in its general
form.

It is possible to introduce a deformation parameter into
the \faa representation of $\Win1\,$ algebra by redefining $W_n$ to
$W_n (h) = (-1)^n \bj (h D + J)^n \cdot 1$ and
accordingly $\O^{(r)}$ \ct{2boson}.
The semiclassical limit is simply obtained by taking $h \to 0$
in $W_n (h)$ and yields the generators of area preserving
${\bf w_{1+\infty}}$ algebra.

The higher bracket structures have been investigated in \ct{BAK85,2boson}.
We write the second bracket in the form:
\br
\{ \bj (x) \, , \, J (y) \}_2 &=& J(x) \d^{\pr} (x-y) - h \d^{\pr\pr} (x-y)
\nonu\\
\{ \bj (x) \, , \, \bj (y) \}_2 &= &  2 \bj (x) \d^{\pr} (x-y) +\bj^{\pr}
(x) \d (x-y) \lab{2jbarj}\\
\{ J (x) \, , \, J (y) \}_2 &=& c \, \d^{\pr} (x-y) \nonu
\er
Although Lenard relations require that the constants
$h$ and $c$ are fixed to $h=1$ and $c=2$ we keep them general
in \rf{2jbarj} in order to be able to interpret them as
independent deformation parameters.
This will simplify our discussion in what follows.
Among the Poisson bracket structures only first and second are
independent, the higher bracket can be given in terms of the first two
by means of the recurrence relations \ct{2boson}.

{\sl Second Bracket from Two-Boson KP System.}
We will now show how to generate the second bracket
structure from the representation given by
\rf{wnpj}.
The algebra of $J$ and $\bj$ will be given by \rf{2jbarj}.
Here we take $h=1$.
Letting again $\Phi$ be such that $\Phi^{\pr}(x) = J (x)$ we derive
{}from \rf{2jbarj}:
\be
\lcurl \bj (x) \, , \, \exp \( \pm \Phi (y)\) \rcurl_2 = \mp \d(x-y)
\pa \exp \( \pm \Phi (y)\) \pm \d^{\pr} (x-y) \exp \( \pm \Phi (y)\)
\lab{2jbarphi}
\ee
repeating similar calculation as in \ct{2boson} we get
for the linear part of $\{\cdot,\cdot\}_2$ the expected
result $\{W_n (x) \, , \, W_m (y) \}_2 \v_{\rm linear} =
\O_{nm}^{(1)} (W(x))\, \d (x-y)$.
To calculate the nonlinear part of the bracket we will use the
exponential representation of \faa polynomials
$P_n (J) = \exp (- \Phi) \pa^n \exp (\Phi)$.
We first observe that
\be
\{ \Phi (x) \, , \, \Phi (y) \}_2 = - c \, \vareps (x-y)
\lab{phiphi}
\ee
from which, the direct calculation yields
\br
\lefteqn{
\lcurl P_n (x) \, ,\, P_m (y) \rcurl_2 = - c
\Bigg\lb \sum_{l=0}^n \sum_{p=0}^m {n \choose l} {m \choose p}
P_{n-l} (x) P_{m-p} (y) \pa^l_x \pa^p_y }\nonu\\
&-& P_{n} (x) \sum_{l=0}^m {m\choose l}
P_{m-l}(y)\pa^l_y -P_{m} (y) \sum_{l=0}^n {n\choose l} P_{n-l}(x)
\pa^l_x \nonu \\
&+&P_{n} (x) P_{m}(y) \Bigg\rb \vareps (x-y)
\lab{nonlinpart}
\er
where we wrote for brevity $P_n \( J(x) \) = P_n (x)$.
In the final expression pure $\vareps (x-y)$ terms
cancel out leaving only delta functions and their derivatives:
\be
\lcurl P_{n} (x) \, ,\, P_{n} (y) \rcurl_2
= -c \Bigg\lb \sum_{l=1}^n \sum_{p=1}^m (-1)^{p} {n \choose l}
{m \choose p} P_{n-l}(x) P_{m-p} (y) \pa^{l+p-1}_x \Bigg\rb \d (x-y)
\lab{pnpmnonlin}
\ee
We obtain therefore the complete second bracket for the generators in
\rf{wnpj} as the sum of linear and nonlinear terms (after
a change of variables $n-l=i$, $m-p=j$ in \rf{pnpmnonlin}):
\br
\lefteqn{
\lcurl W_{n}(x) \, ,\, W_{m} (y) \rcurl_2 = \O_{nm}^{(1)}\(W(x)\)\,
\d (x-y) }\lab{third}\\
&-& c \Bigg\lb \sum_{i=0}^{n-1} \sum_{j=0}^{m-1} (-1)^{n-i}
{n \choose i} {m \choose j} W_{i}(x) D^{n+m-i-j-1}_x W_{j} (x)
\Bigg\rb \d (x-y)   \nonu
\er
for $n,m \geq 0$.

One recognizes in the second term on the right hand side of \rf{third}
only the DS structure from \rf{dscomp} multiplied by $c$ while the
nonlinear part of the second Gelfand-Dickey bracket from
\rf{gd2comp} appears to be missing.
This rises the question whether we have an agreement with the second
Hamiltonian structure of \rf{2bracket} for our basis \rf{wnpj}.
The following identity
\be
\{ W_n (x) \, , \, W_m (y) \}^{GD}_2 \bigg\v_{\rm nonlinear}
= \{ W_n (x) \, , \, W_m (y) \}^{DS}
\lab{2=3}
\ee
valid for the generators $W_n (x) = (-1)^{n}\, \bj (x) \, P_n (J(x))$
shows that this is indeed the case (for the proof see \ct{2boson}).

As a consequence we are able to rewrite relation \rf{third} for
the special case of $c=2$ as
\be
\lcurl W_{n}(x) \, ,\, W_{m} (y) \rcurl_2
= \{ W_n (x) \, , \, W_m (y) \}_2^{GD} + \{ W_n (x) \, , \, W_m (y) \}^{DS}
\lab{w2bracket}
\ee
where on the right hand side of \rf{w2bracket} we had split nonlinear
part of \rf{third} equally between $\{\cdot,\cdot\}_2^{GD} \v_{\rm
nonlinear}$ and $\{\cdot,\cdot\}^{DS}$.
\lskip
{\sl Quadratic Two-Boson KP Hierarchy.}
Here we call quadratic two-boson KP hierarchy
the model based on the pseudo-differential operator \ct{YW9111}:
\be
L_{\sj} = D + \bsj\, \(D - \sj \,- \bsj \,\)^{-1} \sj
\lab{sjlax}
\ee
Let us discuss the Poisson bracket structure first.
Among the Poisson bracket structures only second and third are local.
The non-vanishing bracket within the second structure is given by
\be
\pbr{\sj\,(x)}{\bsj\,(y)}_2 \, =\, \d^{\pr} (x-y)
\lab{p1sjp2sj}
\ee
As shown in \ct{AFGMZ} the Hamiltonian structure corresponding to the
Lax operator $L_{\sj}$ in \rf{sjlax} is invariant under the
transformation $g$ and its inverse $ g^{-1}$
\br
 g\, ( \sj\, ) &=&  \bsj - {\sj^{\,\pr}\o \sj} \qquad, \qquad
g\, ( \bsj\, )\, =\, \sj \lab{gtransf}\\
g^{-1}\, (\bsj\,) &=& \sj\, + { \bsj^{\,\pr} \o \bsj } \qquad, \qquad
g^{-1} \,(\sj\,)\, = \, \bsj \lab{ginvtransf}
\er

Hence we conclude that the Lax operator given by e.g.
\be
L_{\sj} = D + \( \sj + {\bsj^{\, \pr} \o \bsj} \)
\(D - \sj \,- \bsj \,-{\bsj^{\,\pr} \o \bsj} \)^{-1}  \, \bsj
\lab{sjlax2}
\ee
leads to the same Hamiltonian functions as \rf{sjlax}.
\lskip
{\sl Gauge Equivalence between \faa and Quadratic Two-Boson Hierarchies
and Generalized Miura Maps.}
We apply on $L_{\sj}$ from \rf{sjlax} the gauge transformation
generated by $\chi  = \( \p +\bp - \ln \sj\,\)$ with result:
\be
L_{\sj} \to \exp ( - \chi )\,L_{\sj}\,
\exp (\chi ) = D + \sj \,+\bsj \, +
\sj \, ( \sj^{-1} \, )^{\pr} \, +  \bsj \, \sj \,
D^{-1} = D - J + \bj D^{-1}   \lab{lsjgauge}
\lab{gaugea}
\ee
where we have introduced
\be
J = - \sj \,- \bsj \, + {\sj^{\,\pr}  \o \sj}  \qquad ; \qquad
\bj = \bsj \, \sj            \lab{gmiura}
\ee
One can now verify explicitly that with the bracket structure given by
\rf{p1sjp2sj} the variables defined in \rf{gmiura} satisfy
the second bracket structure \rf{2jbarj} of \faa hierarchy.
Hence we obtained a Miura transform for two-Bose hierarchies in
form of \rf{gmiura}, which generalizes the usual Miura transformation
between one-bose KdV and mKdV structures \ct{AFGMZ}.
As a corollary we also obtained in this way a short proof for the quadratic
two-boson KP hierarchy system realizing $\hWinf$.

With the Lax operator from \rf{sjlax2} we find $L_{\sj} =
\bsj^{\,-1} \bsj \, L_{\sj} \bsj^{\,-1} \bsj \,=\, D +
\(\bsj\, \sj + \bsj^{\, \pr} \,\) \(D - \sj \,- \bsj \,\)^{-1}$.
The appropriate gauge transformation gives
\be
L_{\sj} \sim \exp \(- \p - \bp \) \,L_{\sj} \exp \( \p + \bp \)
= D + \sj \,+\bsj \,+\, \(\bsj\, \sj + \bsj^{\, \pr} \,\) D^{-1}
\lab{sjlaxd}
\ee
producing object in \faa hierarchy with
$ J =- \sj \,-\bsj \,$ and $ \bj = \bsj\, \sj \,+\bsj^{\,\pr} \, $.
Note, that under $g^{-1}$ from \rf{ginvtransf} these variables are
transformed into $ J = -\sj \,-\,\bsj \,+ {\sj{\,\pr} \o \sj }\,$ and
$ \bj = \bsj\, \sj \,$ identical to \rf{gmiura}.

We see that because of \rf{gtransf} and \rf{ginvtransf} there is an ambiguity
in the possible form of generalized Miura transformation and \rf{gmiura} can
also appear in other forms. All of them are connecting the Poisson bracket
structure of \faa hierarchy with the corresponding Poisson bracket structure
of the quadratic two-boson hierarchy.
\lskip
{\large {\bf 4. Lagrangian Models and their W-Infinity Invariance}}
\lskip
Given is a generalized Liouville Lagrangian:
\be
{\cal L} = \pa_{+} \p_{-} \pa_{-} \p_{-} -
\pa_{+} \p_{+} \pa_{-} \p_{+} + \kappa e^{2\p_{-}}
\lab{liolag}
\ee
which differs from the one considered in \ct{BI91} by sign in
front of the second term.
We shall take $x_{-}$ to be ``time" and $x_{+}$ to be ``space" variables.
In this setting contribution to the Hamiltonian density comes entirely
from the last term (exponential interaction term).
Therefore the classical Hamiltonian density is bounded no matter what signs
are in front of the kinetic terms.
Lagrangian \rf{liolag} leads to the following
canonical (Dirac) bracket structure:
\be
\{ \p_{\pm} (x) \, , \, \p_{\pm} (y) \} = \mp {1 \o 4} \vareps (x-y)
\lab{liobra}
\ee
which reproduces the second bracket \rf{p1sjp2sj} of the quadratic
two-boson hierarchy when we set:
\be
\sj \, = \pa_{+} \( \p_{+} - \p_{-} \) \quad;\quad
\bsj \, = \pa_{+} \( \p_{+} + \p_{-} \)
\lab{p2sj}
\ee
Furthermore, in this representation we can find via the generalized
Miura transformation given below eq. \rf{sjlaxd}
the  elements $\bj,J$ of the \faa hierarchy as
\br
\bj \eq \bsj \, \sj \,+\bsj^{\,\pr}=  \llbrack (\pa_{+}
\p_{+} )^2 -  (\pa_{-} \p_{-} )^2 \rrbrack + \( \pa_{+}^2 \p_{+} +
\pa_{+}^2 \p_{-} \)   \lab{p2bj} \\
J \eq   - \sj \,-\bsj \, = - 2\, \pa_{+} \p_{+} \lab{p2j}
\er
Recall now that $\bj$ and $J$ are building blocks of the generators
$W_n = (-1)^n \bj P_n (J)$ of $\hWinf$ algebra \rf{2bracket}.

\prop The generalized Liouville Lagrangian \rf{liolag} is invariant
under $\hWinf$ symmetry transformations:
\br
\d^{(n)} \p_{+} \eq \lcurl \int \eps W_n \, , \, \p_{+}\rcurl  =
\sum_{k=0}^{n-1}
{ n \choose k+1} (-1)^{k+n} \pa_{+}^k \( \eps \bj P_{n-k-1} ( J)\)
\lab{dp+}\\
&+& (-1)^n \h \llbrack \eps P_n (J) J + \pa_{+} \( \eps P_n (J) \)
\rrbrack \qquad \;\;\;\;\;\;  n=0,1,2,\ldots  \nonu \\
\d^{(n)} \p_{-} \eq \lcurl \int \eps W_n \, , \, \p_{-} \rcurl  =
(-1)^{n+1} \llbrack \eps P_n (J) \pa_{+} \p_{-}  +
\h \pa_{+} \( \eps P_n (J) \) \rrbrack
\lab{dp-}
\er
with $\pa_{-} \eps =0$.

\proof First note that the interaction term $\int V = \int \kappa e^{\p_{-}}$
is invariant independently of the remaining kinetic terms in
\rf{liolag} as follows from \rf{dp-}:
\be
\d^{(n)} V = 2 \kappa \( \d^{(n)} \p_{-} \)\, e^{2 \p_{-}}= (-1)^{n+1}
\, \kappa\, \pa_{+} \llbrack \eps P_n (J) e^{2\p_{-}} \rrbrack
\lab{dvertex}
\ee
The rest of the proof follows by inspection. It is in fact easy to check
that combination of two kinetic terms (but not single terms separately)
in \rf{liolag} is invariant.
This requires only one technical identity:
\be
\pa_{-} P_n (J) = -2 \sum_{k=0}^{n-1} { n \choose k+1} \( \pa_{+}^{k +1}
\pa_{-} \p_{+} \) P_{n-k-1} ( J)              \lab{bform}
\ee
where we used that from \rf{p2j} we can write the \faa polynomials as
$ P_n (J) = \exp ( \p_{+}) \pa_{+}^n \exp (- \p_{+})$.

Since the symmetry transformations in \rf{dp+} and \rf{dp-} are
generated by $W_n$'s from \rf{wnpj} this establishes invariance of the
generalized Liouville action \rf{liolag} under $\hWinf$ algebra.

As an example of \rf{dp-} and \rf{dp+} let us take the case of $n=0$
for which we find
\br
\d^{(0)} \p_{+} \eq - \eps \, \pa_{+} \p_{+}  +  \h \pa_{+} \eps
\lab{edp+}\\
\d^{(0)} \p_{-} \eq -  \eps  \pa_{+} \p_{-}  - \h \pa_{+} \eps
\lab{edp-}
\er
We note, that $\exp \(2 \p_{\pm}\)$ transform according to \rf{edp+}
and \rf{edp-} as conformal primary fields of weights $\mp 1$,
respectively.
We also note, that $\d^{(n)}$ transformation of $\p_{-}$ can be viewed as
a conformal transformation with the field dependent parameter,
namely $\eps \to \eps P_n (J)$.

We now generalize the previous construction to the setting of Toda
theory associated to a finite Lie algebra $\lie$:
\be
{\cal L} = \h \sum_{a,b}^{r} \eta_{ab} \(
\pa_{+} \p_{-}^a \pa_{-} \p_{-}^b -
\pa_{+} \p_{+}^a \pa_{-} \p_{+}^b \) + \kappa \sum_{a}^{r}
{2 \o \a^2_a} e^{K_{ab} \phi_{-}^b}
\lab{todlag}
\ee
where $K_{ab} = 2 \a_a \cdot \a_b /\a_b^2$ is the Cartan matrix of
$\lie$, furthermore $\eta^{ab} = {\a^2_b \o 2} (K^{-1})^{ab}$,
while $\a_a$ denote the simple roots of $\lie$.
The corresponding brackets are
\be
\pbr{\p_{\pm}^a(x)}{\p_{\pm}^b(y)} = \mp \h \vareps (x-y)\, \eta^{ab}
\lab{todbra}
\ee
Next, we construct two-currents $\bj,J$:
\br
\bj \eq \h  \sum_{a,b}^{r} {2 \o \a^2_a} K_{ab}
\llbrack  \pa_{+} \p_{+}^a \pa_{+} \p_{+}^b
-\pa_{+} \p_{-}^a \pa_{+} \p_{-}^b \rrbrack
+ \sum_{a}^{r} {2 \o \a^2_a}
\llbrack  \pa_{+}^2 \p_{+}^a + \pa_{+}^2 \p_{-}^a \rrbrack
\lab{pa2bj} \\
J \eq   \sum_{a}^{r} \Gamma_a \pa_{+} \p_{+}^a \lab{pa2j}
\er
where we introduced constants $\Gamma_a$ to be determined later.
Construction of $\bj$ is reminiscent of that of E-M tensor and signs are
chosen in such a way that the corresponding Virasoro algebra is
centerless. The currents $\bj,J$ enter the algebra \rf{2jbarj} with
constants
\be
h = - \sum_{a,b}^{r} {2 \o \a^2_a}\, \Gamma_b\, \eta^{ab}
= - \sum_{a}^{r} \Gamma_a\, r^{a}
\quad;\quad c= \sum_{a,b}^{r}\,  \Gamma_{a}\, \Gamma_{b}\, \eta^{ab}
\lab{todhc}
\ee
where $r^a = \sum_{b} \(K^{-1} \)_{ab}$ are the conformal weights of
$\exp (- \p_{-}^a)$ \ct{CFGZ}.
We can choose $\Gamma_a= - K_{ma}$, for fixed but arbitrary $m$,
corresponding to a long simple root $\a_m$ with $\a^2_m=2$ of the
corresponding Lie algebra $\lie$.
In this case, recalling that $\eta^{ab} = (\a^2_b /2)
\(K^{-1} \)_{ab}$, it follows that $h=1$ (with $h$ coinciding
with conformal weights of vertices $\exp \( K_{ab} \p_{-}^b \)$).
Furthermore, it also follows that $c = K_{mm} (\a^2_m /2) =2$.
Hence for an arbitrary Lie algebra $\lie$, currents $\bj, J$ will satisfy
\rf{2jbarj} with $h=1$ and $c = 2$ leading to the $\hWinf$ algebra
of the form \rf{w2bracket}.

For symmetry transformations we find in this case:
\br
\d^{(n)} \p_{+}^c \eq \lcurl \int \eps W_n \, , \, \p_{+}^c \rcurl
= \sum_{k=0}^{n-1} (-1)^{k+n+1} {n \choose k+1} \sum_b^r \, \Gamma_b \,
\eta^{bc}\, \pa_{+}^k \( \eps \bj P_{n-k-1} ( J)\) \lab{toddp+}\\
&+& (-1)^{n+1} \eps \, P_n (J) \pa_{+} \p_{+}^c  + (-1)^n \sum_{a}^{r}
{2 \o \a^2_a}\, \eta^{ac} \,\pa_{+} \( \eps P_n (J) \) \nonu \\
\d^{(n)} \p_{-}^c \eq \lcurl \int \eps W_n \, , \, \p_{-} \rcurl
= (-1)^{n+1} \( \eps \, P_n (J) \pa_{+} \p_{-}^c  +
\sum_{a}^{r} {2 \o \a^2_a}\, \eta^{ac}\, \pa_{+} \( \eps P_n (J) \) \)
\lab{toddp-}
\er
where generators $W_n$ have the same functional form as in \rf{wnpj}
in terms of currents $\bj,J$ from \rf{pa2bj} and \rf{pa2j}.
As before the vertex part is invariant due to:
\be
\d^{(n)} \llbrack \sum_{a}^{r} {2 \o \a^2_a} e^{K_{ab} \phi_{-}^b}
\rrbrack = (-1)^{n+1} \sum_{a}^{r} {2 \o \a^2_a} \llbrack
\eps P_n (J) (\pa_{+} e^{K_{ab} \phi_{-}^b} ) + (\pa_{+} \eps P_n (J) )
e^{K_{ab} \phi_{-}^b} \rrbrack
\lab{dtodvertex}
\ee
being obviously  a total derivative.
Proof that the remaining free part of Langrangian is invariant is a
simple generalization of the proof given above for invariance of the
free part of the generalized Liouville action with the technical
identity \rf{bform} being now replaced by
\be
\pa_{-} P_n (J) = \sum_{a}^r \,\Gamma_a \,\sum_{k=0}^{n-1} { n \choose k+1}
\( \pa_{+}^{k +1} \pa_{-} \p_{+}^a \) P_{n-k-1} ( J) \lab{todbform}
\ee
\lskip
{\large {\bf 5. Discussion }}
\lskip
Several remarks are in order.

Since the symmetry is non-linear the knowledge of exact functional form
of the generators is required to determine all transformations
$\d^{(n)} \p_{\pm} $. This knowledge seems necessary
even if one tries to find higher transformations
of $\p_{\pm}$ just by commuting the lower ones (the argument usually used in
the case of linear symmetry) due to non-linearity.

With the minus sign in front of the second term in \rf{liolag} the field
$\p_{+}$ appears to be a ghost field introduced to enhance the symmetry
structure of the original (conformal) Liouville model.
Any attempt to change this minus sign to a plus sign will either render the
algebra of $\bj$ anomalous (breaking the above symmetry structure)
or will result in introduction of the imaginary constants in the
definition of the symmetry transformations.

The models defined by \rf{liolag} and \rf{todlag} are integrable.
Recall namely the definition of infinitely many charges in involutions
$H_n ( \bj,J)$ associated with two-boson \faa hierarchy (see e.g.
\ct{AFGMZ}).
They all commute with the vertex Hamiltonians of \rf{liolag} and
\rf{todlag} and are therefore conserved.
This result follows from the bracket relations:
\be
\pbr{J(x)}{V (y)} = 0  \quad;\quad
\pbr{\bj(x)}{V (y)} = V (x) \d^{\pr} (x - y)
\lab{Jvertex}
\ee
which hold both for the vertex $V (y) =e^{2 \p_{-}(y)}$ of the generalized
Liouville model as well as for the vertices $V (y) =
\kappa \sum_{a}^{r} {2 \o \a^2_a} e^{K_{ab} \phi_{-}^b (y)}$
of the generalized Toda model.
Hence both models are integrable and the underlying structure
of the two-boson KP hierarchy allows for explicit construction
of charges.

We have seen that the symmetry transformations: $ g ( \sj \,) = \bsj \,
- ( \ln \sj \, )^{\pr} \, , \, g (\bsj \,) = \sj \, $ as well as its
inverse  $g^{-1}$ keep all Hamiltonians invariant.
By acting with $g$ and $g^{-1}$ and their powers one obtains the family
of vertices: $\sj \, \exp (-2 \p_{-})$, $\bsj \, \exp (-2 \p_{-}),\ldots$
etc.
They will all commute with charges $H_n $.
Moreover it is possible to associate to each of these vertices a
Lagrangian density invariant under $\hWinf$ but with kinetic
terms containing $\p_{\pm}$ transformed by the corresponding $g$
transformations.
One obtains in this process instead of the single Lagrangian \rf{liolag}
a string of Liouville like actions, which reflect existence of the
symmetry $g$ in the underlying two-boson KP hierarchy.
For instance acting once with $g$ on Lagrangian in \rf{liolag}
leads to:
\be
{\cal L} = \pa_{+} \p_{-} \pa_{-} \p_{-} -
\pa_{+} \p_{+} \pa_{-} \p_{+} +
{ \pa_{+} \lb \pa_{+} \( \p_{+} - \p_{-} \)
\pa_{-} \( \p_{+} - \p_{-} \) \rb \o 2 \pa_{+} \( \p_{+} - \p_{-} \)}
+ \kappa  \, \pa_{+} \( \p_{+} - \p_{-} \)   e^{-2 \p_{-}}
\lab{gliolag}
\ee
which also is $\hWinf$ invariant with $\bj$ and $J$ realized as in
\rf{gmiura}.

As it was shown in \ct{AFGMZ} taking the Dirac constraint $ J=0$ reduces
the two-boson hierarchies to KdV and mKdV systems.
Correspondingly removing $ \p_{+} $ in \rf{liolag} casts the Lagrangian
to the form of the usual Liouville model. While the vertex
is still commuting with Hamiltonians of the reduced systems (KdV)
the corresponding symmetry structure shrinks to
the conformal symmetry generated by the reduced current
$\bj = - \sj^{\,2} + \sj^{\,\pr}$.
\lskip
{\bf Acknowledgements}
We gratefully acknowledge support within CNPq/NSF Cooperative Science Program.
One of us (HA) thanks Instituto de F\'{\i}sica Te\'{o}rica-UNESP
for kind hospitality.

\end{document}